\documentclass[a4paper]{revtex4}

\usepackage{amsmath,amssymb,amsfonts,amstext}
\usepackage{setspace}                

\usepackage{graphicx,color}          
\usepackage{epsfig}                  
\usepackage{wrapfig}
\usepackage{subfigure}
\usepackage{longtable}               
\usepackage{mathrsfs}                

\usepackage{amsthm}
\theoremstyle{plain}

\usepackage{dsfont}                  
\usepackage[a4paper]{anysize}        
\usepackage{braket}

\newtheorem{lemma}{Lemma}

\newtheorem{theorem}{Theorem}

\newtheorem{cor}{Example}

\newtheorem*{LB}{Lieb-Robinson Bounds}

\begin{document}

\title{Bounding entanglement spreading after a local quench}

\author{Raphael C. Drumond}
\affiliation{Departamento de Matem\'{a}tica, Instituto
de Ci\^{e}ncias Exatas, Universidade Federal de Minas Gerais, CP
702, CEP 30123-970, Belo Horizonte, Minas Gerais, Brazil.}%
%
\author{Nat\'{a}lia S. M\'{o}ller}

\affiliation{Departamento de F\'isica, Instituto
de Ci\^{e}ncias Exatas, Universidade Federal de Minas Gerais, CP
702, CEP 30123-970, Belo Horizonte, Minas Gerais, Brazil.}

\begin{abstract}

We consider the variation of von Neumann entropy of subsystem reduced states of general many-body 
lattice spin systems due to local quantum quenches. We obtain Lieb-Robinson-like bounds that 
are independent of the subsystem volume. The main assumptions are that the Hamiltonian satisfies a 
Lieb-Robinson bound and that the volume of spheres on the lattice grows at most exponentially with 
their radius. More specifically, the bound exponentially 
increases with time but exponentially decreases with the distance between the subsystem and the 
region where the quench takes place. The fact that the bound is independent of the subsystem volume 
leads to stronger constraints (than previously known) on the propagation of information 
throughout many-body systems. In particular, it shows that bipartite entanglement satisfies an 
effective ``light cone,'' regardless of system size. Further 
implications to $t$ density-matrix renormalization-group simulations of quantum spin chains and limitations to the propagation of information are discussed. 
\end{abstract}


\maketitle

\section{Introduction}

Entanglement is a fundamental quantity of quantum information and 
computation, being essential to perform tasks such as teleportation or superdense 
coding~\cite{quantcomp}. In recent years it is becoming increasingly relevant also 
to quantum many-body physics. It can be a good order parameter for quantum 
phase transitions~\cite{brandao}. Algorithms for computing one-dimensional 
quantum many body ground states, such as the density matrix 
renormalization group (DMRG)~\cite{dmrg} method or the variational calculus over matrix 
product 
states (MPS)~\cite{mps}, have their 
efficiency based essentially on the spatial scaling of entanglement within these 
states~\cite{dmrgmps}. It is a key ingredient for the (subsystem) thermalization 
of many-body isolated quantum systems~\cite{therm}.

Entanglement may also be of interest for non-equilibrium phenomena~\cite{global, 
local}. The spatial scaling of entanglement within the eigenstates of a many-body Hamiltonian, as 
well as its growth in time, is a signature of the many-body localized phase~\cite{mblestat}. 
The dynamics of entanglement due to global or local quenches may 
be computed by conformal field theory techniques~\cite{cft}, or by the time variants of DMRG or MPS based 
algorithms~\cite{tdmrg}, or at least have its growth bounded~\cite{Bravyi,bounded}. 

The behavior of a many-body system after a quantum quench can raise 
fundamental questions, such as whether the system equilibrates or not (see, 
e.g.,~\cite{equi}). It can be investigated with 
increasing detail in modern experimental settings such as ultracold atoms in optical 
lattices~\cite{optlat} or trapped 
ions~\cite{ion}. Moreover, novel numerical techniques, such as $t-$DMRG, allow one to simulate the 
evolution of significantly large systems, especially spin chains~\cite{tdmrg}. In simulations of 
quantum chains by $t-$DMRG the entanglement of every bipartition of the chain (in two 
contiguous regions) is naturally computed for every instant of time. After a local quantum quench, 
it can be seen, for instance in Ref.~\cite{alberto}, that entanglement of these 
bipartitions satisfies an effective ``light cone'' in the same way as any other local quantity of the 
system, such as magnetization.  

In \cite{Bravyi} this light cone effect can be partially explained for a 
local quench on the initial state of the system. There, a unitary operation with support on a small 
region of the system can be applied, with the purpose of 
establishing a communication channel between distant regions of the system. The authors of \cite{Bravyi} estimate the 
variation of quantum entropy---with respect to the evolutions with and without an applied 
unitary---for any region away from the quench. They found a bound for its growth in time assuming a 
Lieb-Robinson bound~\cite{LR} for the model. However, their bound is proportional to 
the volume of the region, restricting its validity. For instance, it can not be applied if 
one takes the thermodynamic limit of the subsystem. Moreover, the bound could not be used to 
guarantee an area law for entanglement~\cite{arealaw} of the evolved states, since it is 
proportional to the subsystem volume.

Here we provide Lieb-Robinson-like bounds for the variation of quantum entropy of the reduced 
states of any region away from a quench. We 
consider two kinds of quenches: a local perturbation on the Hamiltonian and on the initial state. We 
assume only that the model satisfies a Lieb-Robinson bound and that the volume of lattice spheres 
grows at most exponentially with their radius. 

We discuss three consequences of the 
bounds. First, we show the validity of an 
effective light cone for entanglement, in a sense we shall explain in detail later. 
Second, we point out how the 
bounds guarantee, for every instant of time, an area law for entanglement of the evolved states, 
as long as the initial state also satisfies an area law and is an eigenstate of the Hamiltonian. 
And third, we discuss how the bound implies a strong restriction on the information capacity of 
quantum channels established between distant regions of a many-body system.

This paper is organized as follows. In Sec.~\ref{Setting} we define the class of models we 
shall deal with, and we state a Lieb-Robinson bound and further necessary concepts and results. In 
Sec.~\ref{results} we prove bounds for the variation of entanglement after a local quench and 
point out some special cases. In Sec.~\ref{conclusion} 
we discuss some implications of the bounds obtained. 

\section{Preliminaries}\label{Setting}

Schr\"odinger's equation is non-relativistic, so, in principle, it does not 
forbid instantaneous propagation of information across space. On the other hand, the seminal paper 
by Lieb and Robinson \cite{LR} suggests that a \emph{de facto} causality should be valid when 
a perturbation propagates on a many-body system with short-range interactions. Further 
refinements~\cite{Sims} of their work led to a number of results, collectively known as 
Lieb-Robinson bounds. In Ref.~\cite{Bravyi} the authors show that if a many-body 
system satisfies a Lieb-Robinson bound, there is indeed a limit for the speed of propagation of 
(any significant amount of) information. In the following we shall recall the large 
class of quantum many-body systems considered in Ref.~\cite{Sims} 
for which the authors derive Lieb-Robinson bounds. 

\subsection{Lieb-Robinson Bounds}\label{LRsection}

A quantum many-body spin model is given by a triple $(\Gamma,\{\mathcal{H}_{i}\}_{i\in 
\Gamma},\Phi)$ where $\Gamma$ is a metric space, $\mathcal{H}_{i}$ is a Hilbert space for every 
$i\in\Gamma$, and $\Phi$ is an interaction. We shall assume for simplicity that $\Gamma$ is the set 
of vertices of a connected graph, imbued with the set-theoretical distance. Namely, for every 
$i,j,\in\Gamma$, 
the distance $d(i,j)$ between them is the length of a shortest path connecting $i$ and $j$. Each 
point $i$ of $\Gamma$ describes an 
individual quantum system with finite dimensional Hilbert space $\mathcal{H}_{i}$. For any finite 
subset $\Lambda$ of $\Gamma$ the corresponding state space is 
$\mathcal{H}_{\Lambda}=\bigotimes_{i\in \Lambda}\mathcal{H}_{i}$. The interaction 
$\Phi$ associates to every finite subset $X$ of $\Gamma$ a self-adjoint operator $\Phi(X)$ on 
$\mathcal{H}_X$. Finally, for every finite $\Lambda\subset \Gamma$ the Hamiltonian of that 
portion of the system is defined by $H_{\Lambda}=\sum_{X\subset\Lambda} \Phi(X)\otimes 
\mathds{1}_{\Lambda\backslash X}$.

In order to get a Lieb-Robinson bound, the interaction must decay fast enough with the diameter of 
finite subsets of $\Gamma$. This is encoded by a non-increasing function 
$F:=[0,\infty)\rightarrow 
(0,\infty)$ that must satisfy, for every $\mu\geq 0$: $$||F||:=\sup_{i\in \Gamma}\sum_{j\in 
\Gamma}F(d(i,j))<\infty,$$ $$C_{\mu}:=\sup_{i,j\in \Gamma}\sum_{k\in 
\Gamma}\frac{e^{-\mu[d(i,k)+d(k,j)-d(i,j)]}F(d(i,k))F(d(k,j))}{F(d(i,j))}<\infty,$$ where $d(i,j)$ is 
the distance between $i,j\in \Gamma$. With such an $F$, the following condition guarantees a fast 
enough decay of $\Phi$: $$||\Phi||_{\mu}:=\sup_{i,j \in \Gamma}\sum_{X \ni 
i,j}\frac{||\Phi(X)||}{e^{-\mu(d(i,j))}F(d(i,j))}<\infty.$$ Defining the $\Phi$ boundary of a subset 
 $X$ by $\partial_{\Phi}X=\{i\in X: \exists Y\subset\Gamma \text{ with } Y\cap 
X^c\neq\varnothing, i\in Y \text{ and } \Phi(Y)\neq 0\}$, and by $|X|$ the number of elements 
of a set $X$, the following bound can then be obtained~\cite{Sims}:
\begin{LB} Let $X,Y\subset \Lambda$ with $d(X,Y)>0$; let $A$ and $B$ operators be defined on $H_\Lambda$ 
with support on $X$ and $Y$, respectively; and let $A(t)=e^{iH_{\Lambda}t}Ae^{-iH_{\Lambda}t}$. Then, 
the following inequality holds true for every $\mu>0$  and $t \in \mathds{R}$:
\begin{align}||[A(t),B]||\leq\frac{2||A||||B||||F||}{C_{\mu}}\min{\{|\partial_{\Phi}X|,|\partial_{
\Phi}Y|\}}e^{-\mu(d(X,Y)-v_{\mu}|t|)},\label{LiebRob}
\end{align}
where $v_{\mu}=\frac{2||\Phi||_{\mu}C_{\mu}}{\mu}$.
\end{LB}

\subsection{Continuity Inequalities for Entropy}

For estimating the variation of entanglement we shall need to bound the variation of reduced 
states of the system, measured by the trace distance, as well as continuity inequalities for 
quantum entropy. 

Let the trace norm of an operator $A$ be given by
\begin{equation}
 ||A||_{1}=\sup_{||U||=1}\{|\text{Tr}AU|\} \label{tracenorm}
\end{equation}
and let $S(\rho)=-\text{Tr}(\rho\log_2{\rho})$ be the von Neumann entropy of a density operator 
$\rho$ acting on a Hilbert space of dimension $D$. The following continuity inequality holds~\cite{Audenaert}:
\begin{gather}
|S(\rho)-S(\rho')|\leq \frac{1}{2}||\rho-\rho'||_{1}\log_2{(D-1)}+h(\frac{1}{2}||\rho-\rho'||_{1})\label{auden},
\end{gather}
where $h(x)=-x\log_2{x}-(1-x)\log_2{(1-x)}$ is the binary entropy function. 
Moreover, the quantum conditional entropy 
$S_{X|Y}(\rho_{XY})=S(\rho_{XY})-S(\rho_{Y})$, where $\rho_{XY}$ is a state of a bipartite system 
$XY$ and $\rho_{Y}$ is the corresponding reduced state of part $Y$, satisfies the continuity 
inequality~\cite{alicki}: 
\begin{equation}
|S_{X|Y}(\rho_{XY})-S_{X|Y}(\rho'_{XY})|\leq 
4||\rho_{XY}-\rho'_{XY}||_{1}\log_2{D_{X}}+2h(||\rho_{XY}-\rho'_{XY}||_{1})\label{ali},
\end{equation} 
valid whenever $||\rho_{XY}-\rho'_{XY}||_{1}<1$, where $D_{X}$ is the Hilbert-space dimension of 
part $X$.

\section{A bound for the variation of von Neumann entropy under local quenches}\label{results}

To understand the spreading of correlations and transport on many-body systems one
may resort, both theoretically and experimentally~\cite{optlat,local}, to
following the dynamics of the system after a local quench. One can distinguish two kinds of local
(instantaneous) quenches: a sudden local change on the Hamiltonian $H$ and on the initial state
$\ket{\psi}$ of the many-body system. That is, in the first case, from time $t=0$ and on, the
Hamiltonian changes to $H+W$. For the second case, an initial state $\ket{\psi}$ is quickly changed 
to
$U\ket{\psi}$. Both $W$ and $U$ must have support on a small portion of the system. In either case,
we can compare the evolution of the system with and without the applied quench.

First we show that, for small times, the reduced state of regions far from the region where the
quench takes place is slightly perturbed. Note that inequality~\eqref{rapbound} shown 
below corresponds for $q=1$ it corresponds to a 
quenched Hamiltonian while for $q=2$ to a quenched initial state.
\begin{lemma}\label{lemma}
Let $(\Gamma,\{\mathcal{H}\}_{i\in\Gamma},\Phi)$ be a model satisfying the conditions 
described in Sec.~\ref{LRsection} and let $\Lambda$ be any finite subset of $\Gamma$. 
Let $X,Y\subset 
\Lambda$ be two subsets with $d(X,Y)>0$. Let $W$ be a self-adjoint operator on 
$\mathcal{H}_{\Lambda}$ and let $U_{X}$ be a 
unitary operator, both of them with support on $X$. Let $\ket{\psi}$ be a unit vector of
$\mathcal{H}_{\Lambda}$ and denote $\ket{\psi^{0}(t)}=e^{-iH_{\Lambda}t}\ket{\psi}$,
$\ket{\psi^{1}(t)}=e^{-i(H_{\Lambda}+W)t}\ket{\psi}$, 
$\ket{\psi^{2}(t)}=e^{-iH_{\Lambda}t}U_{X}\ket{\psi}$. Denote the reduced states on region $Y$ as 
follows 
$\rho^{q}_{Y}(t)=\text{\normalfont{Tr}}_{\Lambda\backslash Y}(\ket{\psi^{q}(t)}\bra{\psi^{q}(t)})$, for 
$q=0,1,2$, are their respective reduced states on region $Y$. For any $\mu > 0$ and $t\in
\mathds{R}$ the following inequality holds true:
\begin{align}||\rho_{Y}^{0}(t)-\rho^{q}_{Y}(t)||_{1}\leq c_{q}e^{-\mu(d(X,Y)-v_\mu 
|t|)}\label{rapbound}
\end{align}
for $q=1,2$, where $c_{1}=\frac{2||W|||F||}{\mu 
v_{\mu}C_{\mu}}\min{\{|\partial_{\Phi}X|,|\partial_{\Phi}Y|\}}$ and 
$c_{2}=\frac{2||F||}{C_{\mu}}\min{\{|\partial_{\Phi}X|,|\partial_{\Phi}Y|\}}$. 
\end{lemma}
\emph{Proof.} First we show the inequality for $q=1$. Let $U_{Y}$ be an operator acting on 
$\mathcal{H}_{\Lambda}$ with support on $Y$ and let $\tilde{U}_{Y}$ be its restriction to 
$\mathcal{H}_{Y}$. We have then~\cite{kastner}:
\begin{align}
|\text{\normalfont{Tr}}\{[\rho^{0}_{Y}(t)-\rho^{1}_{Y}(t)]\tilde{U}_{Y}\}|&=|\bra{\psi^{0}(t)}U_{Y}
\ket {\psi^{0}(t)}-\bra{\psi^{1}(t)}
U_{Y}\ket{\psi^{1}(t)}|\\
&=|\bra{\psi}e^{iH_{\Lambda}t}U_{Y}e^{-iH_{\Lambda}t}-e^{i(H_{\Lambda}+W)t}U_{Y}e^{-i(H_{\Lambda}
+W)t} \ket{\psi}|\\
&\leq 
||e^{iH_{\Lambda}t}U_{Y}e^{-iH_{\Lambda}t}-e^{i(H_{\Lambda}+W)t}U_{Y}e^{-i(H_{\Lambda}+W)t}||\\
&=||e^{-i(H_{\Lambda}+W)t}e^{iH_{\Lambda}t}U_{Y}e^{-iH_{\Lambda}t}e^{i(H_{\Lambda}+W)t}-U_{Y}
||.\\
&=||\int_{0}^{t}dt'\frac{d}{dt'}e^{-i(H_{\Lambda}+W)t'}e^{iH_{\Lambda}t'}U_{Y}e^{-iH_{\Lambda}t'}e^{
i(H_
{ \Lambda } +W)t' } ||\\
&=||\int_{0}^{t}dt'e^{-i(H_{\Lambda}+W)t'}[H_{\Lambda}+W-H_{\Lambda},U_{Y}(t')]e^{-i(H_{\Lambda}+W)t'
}
||\\
&\leq |\int_{0}^{t}dt'||[W,U_{Y}(t')]|||,
\end{align}
where $U_{Y}(t)=e^{iH_{\Lambda}t}U_{Y}e^{-iH_{\Lambda}t}$. Recalling that $W$ 
has support on $X$, we can apply inequality \eqref{LiebRob} to the integrand of the last expression 
and get: $$|\text{\normalfont{Tr}}\{[\rho^{0}_{Y}(t)-\rho^{1}_{Y}(t)]\tilde{U}_{Y}\}|\leq
\frac{2||W|||F||}{C_{\mu}}\min{\{|\partial_{\Phi}X|,|\partial_{\Phi}Y|\}}e^{-\mu d(X,Y)} 
\int_{0}^{|t|}e^{\mu v_{\mu}t'}dt'.$$
Finally, from trace norm characterization \eqref{tracenorm} and observing that 
$\int_{0}^{|t|}e^{\mu v_{\mu}t'}dt'\leq (\mu v_{\mu})^{-1}e^{\mu v_{\mu}|t|}$ we get
inequality~\eqref{rapbound}. 

For $q=2$, take $U_{Y}$, $\tilde{U}_{Y}$ and $U_{Y}(t)$ as above, so~\cite{Bravyi}:
\begin{align}
|\text{\normalfont{Tr}}\{[\rho^{0}_{Y}(t)-\rho^{2}_{Y}(t)]\tilde{U}_{Y}\}|&=|\bra{\psi^{0}(t)}U_{Y}
\ket {\psi^{0}(t)}-\bra{\psi^{2}(t)}
U_{Y}\ket{\psi^{2}(t)}|\\
&=|\bra{\psi}(U_{Y}(t)-U_{X}^{*}U_{Y}(t)U_{X})\ket{\psi}|\\
&\leq ||U_{Y}(t)-U_{X}^{*}U_{Y}(t)U_{X}||\\
&=||U_{X}U_{Y}(t)-U_{Y}(t)U_{X}||\\
&=||[U_{X},U_Y(t)]||.
\end{align}
Again, using the Lieb-Robinson bound~\eqref{LiebRob} and  
expression \eqref{tracenorm} for the trace norm, we get inequality~\eqref{rapbound} 
for $i=2.$
\qed

If we use Lemma~\ref{lemma} above directly with the continuity inequality~\eqref{auden} 
for entropy we obtain bounds for the variation of entropy that grow linearly with $|Y|$, since the 
right hand side of Eq.~\eqref{auden} grows logarithmically with dim$(\mathcal{H}_{Y})$. In order to avoid this we 
can stratify $Y$ in sets of increasing distance to $X$ and compute the entropy as a sum of 
conditional entropies between these sets. The advantage is twofold: (i) the conditional entropies are 
computed on regions of increasing distance to $X$ and, hence, of exponentially decreasing variation; 
(ii) the continuity inequality~\eqref{ali} for conditional entropy depends on the dimension of just one 
of the parts. We must assume, however, that the volume of each set does not grow too fast with its 
distance to $X$ in order to get the desired bound. We shall detail these conditions in the 
following.

For $l\in\mathds{N}$, let $X_{l}=\{j\in\Gamma|d(j,X)=l\}$ be the set of all points of 
$\Gamma$ with distance $l$ to $X$. For $i\in\Gamma$ and $l\in\mathds{N}$, let 
$R_{l}(i)=\{j\in 
\Gamma|d(i,j)=l\}$ be a sphere of radius $l$ centered in $i$. Denote by Int$(X)=\{i\in X 
|R_{1}(i)\subseteq X\}$ the interior of $X$ and let $\partial X=X-\text{Int}(X)$ be its boundary. Note 
that for systems with (non-zero) nearest-neighbor interactions it holds 
that $\partial X=\partial_{\Phi}X$. We must have then the following.
\begin{lemma} \label{lemma2}For every finite $X\subseteq\Gamma$ and $l>0$ it must hold that
\begin{equation}
 X_{l}\subseteq \bigcup_{i\in \partial X}R_{l}(i).
\end{equation}
\end{lemma}
\emph{Proof.} Indeed, take $j\in X_{l}$ and $i\in X$ such that $d(j,X)=d(j,i)=l$. Clearly we have 
$j\in R_{l}(i)$. Take a path of length $l$ 
connecting 
$j$ to $i$. Since $l>0$ this path necessarily contains a point 
$k$ of $R_{1}(i)$. It must hold that $k\notin X$, otherwise one can construct a path of length 
$l-1$ connecting $j$ to a point of $X$, in contradiction with condition $d(j,X)=l$. In other words, 
$i\in \partial X$. \qed

Now we are ready to state the following.
\begin{theorem}\label{thm}
Assume the same conditions and notation of Lemma~\ref{lemma}. Furthermore, assume 
that 
\begin{equation}
|R_{l}(i)|\leq be^{\alpha l} \label{prop}
\end{equation}
for every $i\in \Gamma, l\geq 
0$ and some constants $b,\alpha \geq 
0$. Suppose also that $D=\sup_{i\in\Gamma}\text{\normalfont{dim}}(\mathcal{H}_{i})<\infty$. Let 
$t\in \mathds{R}$ be such that $d(X,Y)>\frac{\mu}{\mu-\alpha}v_{\mu}|t|$. Then, the 
following inequalities hold true:
\begin{equation}
|S(\rho^{0}_{Y}(t))-S(\rho^{q}_{Y}(t))|\leq \gamma_{q}e^{-\frac{\mu}{2}(d(X,Y)-v_{\mu}'|t|)}, 
\label{cota}
\end{equation}
for $q=1,2$ and $\mu>2\alpha$, where
$\gamma_{q}=4\sqrt{c_{q}}(1-e^{-\frac{\mu}{2}})^{-1}(|\partial X|\sqrt{c_{q}}b\log_2{D}+1)$ and 
$v_{\mu}'=\frac{\mu}{\mu-\alpha}v_{\mu}$.
\end{theorem}

\begin{figure}[h]
\includegraphics[scale=0.5]{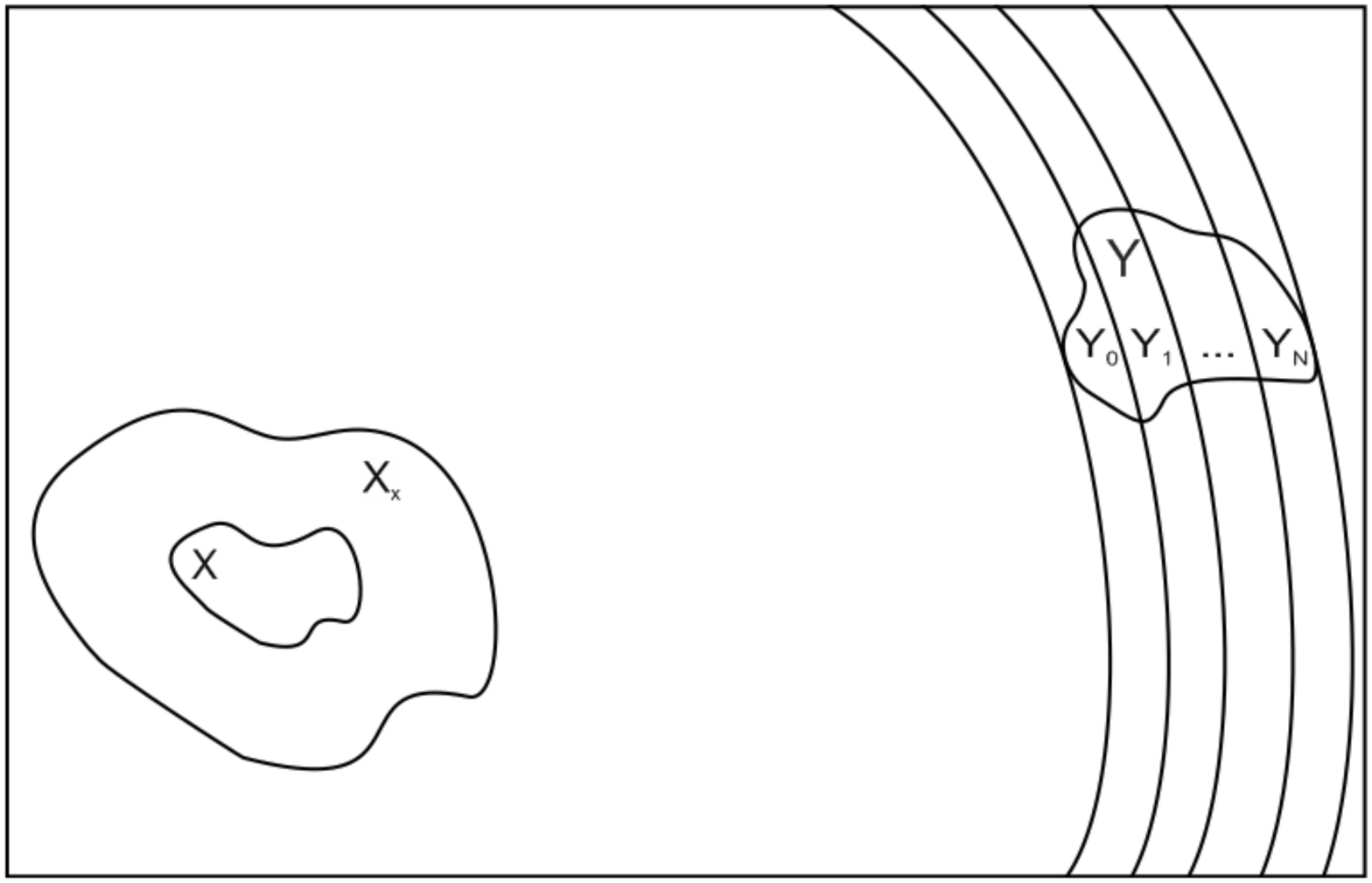} 
\caption{Pictorial depiction of sets $X,Y, Y_{l}$, and $\tilde{X}_{x}$ defined in the 
proof of Theorem~\ref{thm} and in Sec.~\ref{conclusion}.}\label{fig1}
\end{figure}

\emph{Proof.} Define $Y_l=Y\cap X_{d(X,Y)+l}$ for $l\in\mathds{N}$. If 
$N=\max \{l:Y_l\neq\varnothing\}$, the definitions of $N$ and $Y_{l}$ guarantee that 
$Y=\bigcup_{l=0}^NY_l$. Moreover, if 
$\tilde{Y}_l=\bigcup_{m=l}^NY_m$, 
we have $\tilde{Y}_{0}=Y$, $\tilde{Y}_{l}=Y_{l}\bigcup \tilde{Y}_{l+1}$ and 
$Y_{N}=\tilde{Y}_{N}$. See Figure~\ref{fig1} for a pictorial description of all these sets.

All these definitions imply that for any density operator acting on $\mathcal{H}_{Y}$ it 
must hold that:
\begin{equation}
S(\rho_{Y})=\left( 
\sum_{l=0}^{N-1}S_{Y_{l}|\tilde{Y}_{l+1}}(\rho_{\tilde{Y}_{l}})\right)+S(\rho_{Y_{N}}),
\end{equation}
\noindent where $S_{Y_{l}|\tilde{Y}_{l+1}}(\rho_{\tilde{Y}_{l}})$ denotes the conditional entropy
$S_{Y_{l}|\tilde{Y}_{l+1}}(\rho_{\tilde{Y}_{l}})=S(\rho_{\tilde{Y}_{l}})-S(\rho_{\tilde{Y}_{l+1}}
)$. Letting $\Delta S_{q}(t)=S(\rho_{Y}^{0}(t))-S(\rho^{q}_{Y}(t))$ for $q=1,2$, we have on the one 
hand:
\begin{align}
|\Delta S_{q}(t)|&\leq \sum_{l=1}^{N-1}
|S_{Y_{l}|\tilde{Y}_{l+1}}(\rho^{0}_{\tilde{Y}_{l}}(t))-S_{Y_{l}|\tilde{Y}_{l+1}}(\rho^{q}_{\tilde{Y
} _ { l } } (t))| \nonumber
\\&+|S(\rho^{0}_{Y_{N}} (t))-S(\rho^{q}_{Y_{N}}(t))|. \label{entrorelat}
\end{align}
On the other hand, from Lemma~\ref{lemma} we get, for 
$l=0,...,N$: 
\begin{equation}
||\rho_{\tilde{Y}_{l}}^{0}(t)-\rho^{q}_{\tilde{Y}_{l}}(t)||\leq 
c_{q}e^{-\mu(d(X,Y)+l-v_{\mu}t)},\label{ytil}
\end{equation}
since $d(X,\tilde{Y}_{l})=d(X,Y_{l})=d(X,Y)+l$. 
Moreover, by using dim$(\mathcal{H}_{Y_{l}})\leq D^{|Y_{l}|}$, inequalities~\eqref{ytil}, the 
continuity inequalities for entropy \eqref{auden}, and conditional 
entropy \eqref{ali}, the right hand side of Eq.~\eqref{entrorelat} can be bounded by:
\begin{align}
&\sum_{l=0}^{N-1} \{4c_{q} e^{-\mu 
(d(X,Y)+l-v_\mu|t|)}\log_2{(D^{|Y_l|})}+2h(c_{q}e^{-\mu (d(X,Y)+l-v_\mu|t|)})\} \nonumber
\\&+\frac{1}{2}c_{q} e^{-\mu (d(X,Y)+N-v_\mu|t|)}\log_2{(D^{|Y_N|}-1)}+h\left(\frac{1}{2}c_{q} 
e^{-\mu 
(d(X,Y)+N-v_\mu|t|)}\right).
\end{align}
In order to bound the binary entropy functions we use that $h(x)\leq 2\sqrt{x}$ for 
$x\in[0,1]$, so we can 
write:
\begin{align}
|\Delta S_{q}(t)|&\leq\sum_{l=0}^{N-1} \{4c_{q} e^{-\mu 
(d(X,Y)+l-v_\mu|t|)}\log_2{(D^{|Y_l|})}+4\sqrt{c_{q}}e^{-\mu (d(X,Y)+l-v_\mu|t|)/2}\} \nonumber
\\&+\frac{1}{2}c_{q} e^{-\mu (d(X,Y)+N-v_\mu|t|)}\log_2{(D^{|Y_N|}-1)}+\sqrt{2}\sqrt{c_{q}} 
e^{-\mu (d(X,Y)+N-v_\mu|t|)/2}.
\end{align}
In this expression the last two terms are smaller than the term of index $N$ of 
the summand. Therefore, we can bound the expression by a single sum ranging from $0$ to $N$ and get:
\begin{align}
|\Delta S_{q}(t)| &\leq \sum_{l=0}^{N} \{4c_{q} e^{-\mu 
(d(X,Y)+l-v_\mu|t|)}\log_2{(D^{|Y_l|})}+4\sqrt{c_{q}}e^{-\mu (d(X,Y)+l-v_\mu|t|)/2}\}.\label{equu} \\
&= 4c_{q}\log_2{(D)}e^{-\mu (d(X,Y)-v_\mu|t|)}\sum_{l=0}^{N} |Y_l| e^{-\mu l}+4\sqrt{c_{q}}e^{-\mu 
(d(X,Y)-v_\mu|t|)/2}\sum_{l=0}^{N}e^{-\mu l/2},\label{eq2}
\end{align}
where we get the equality by rearranging the terms. We can bound the second summand in 
Eq.\eqref{eq2} immediately by $\sum_{l=0}^{\infty}e^{-\frac{\mu}{2}l}=(1-e^{-\frac{\mu}{2}})^{-1}$. 
To bound the first summand we just have to observe that $|Y_{l}|\leq 
|\bigcup_{i\in \partial X}R_{d(X,Y)+l}(i)|\leq 
|\partial X|be^{\alpha(d(X,Y)+l)}$, where the first inequality comes from the definition of $Y_{l}$ 
and Lemma~\ref{lemma2} while the second comes
from hypothesis~\eqref{prop}. Therefore,
$$\sum_{l=0}^{\infty}|Y_{l}|e^{-\mu l}\leq
|\partial X|be^{\alpha 
d(X,Y)}\sum_{l=0}^{\infty}e^{-(\mu-\alpha)l}=|\partial X|\frac{be^{\alpha 
d(X,Y)}}{1-e^{-(\mu-\alpha)}},$$
since $\mu>\alpha$, and we get
\begin{align}
|\Delta S_{q}(t)|& \leq 4c_{q}|\partial 
X|\log_2{(D)}b(1-e^{-(\mu-\alpha)})^{-1}e^{-(\mu-\alpha)d(X,Y)+
 \mu v_\mu|t|}\\&+4(1-e^{-\frac{\mu}{2}})^{-1}\sqrt{c_{q}}e^{-\frac{\mu}{2}(d(X,Y)-v_\mu|t|)}. 
\label{eq3}
\end{align}
Finally, defining $v_{\mu}'=\frac{\mu}{\mu-\alpha}v_{\mu}$, using that 
$d(X,Y)>v_{\mu}'|t|$ and 
$\mu-\alpha>\frac{\mu}{2}$, we can conclude the desired bound: 
\begin{align}
|\Delta S_{q}(t)|&\leq 
4(1-e^{-\frac{\mu}{2}})^{-1}(c_{q}|\partial X|b\log_2{D}+\sqrt{c_{q}})e^{
-\frac{\mu}{2}(d(X,Y)-v_\mu'|t|)}.
\end{align}
\qed \\

Let us now consider some examples. 

\begin{cor} \label{chain}
If $\Gamma=\mathbb{Z}$ with $d(i,j)=|i-j|$ in Theorem~\ref{thm}, inequality~(\ref{cota}) holds with 
$v_{\mu'}=v_{\mu}$ and $b=2$. Indeed, one just has to realize that such metric space 
$\Gamma$ 
satisfies~\eqref{prop} with $b=2$ and $\alpha=0$ since every sphere in this space has 
precisely two elements, irrespective the size of its radius. 

Assuming further that $r$ is the range of interaction [meaning that 
$\Phi(Z)=0$ for every set $Z$ with diameter larger than $r$]and $X$ is a contiguous region, one has 
$|\partial X|=2$ and $|\partial_{\Phi}X|\leq 2r$. Therefore, the bound is completely independent of 
the size of regions $Y$ and $X$.
\end{cor}

One says that $\Gamma$ has \textit{fractal 
dimension} $n$ if there exists 
$n\geq1$ and $a>0$ such that
\begin{equation}
|R_l(i)|\leq al^{n-1}\label{specdim}
\end{equation}
for every $l>0$~\cite{ref5,refref5}. Note that lattices $\mathbb{Z}^n$ are particular cases of 
such space. In such models one has the following.
\begin{cor}\label{dimn}
In Theorem~\ref{thm}, if $\Gamma$ has fractal dimension $n$, inequality~\eqref{cota} holds for 
every $\alpha>0$ (and $\alpha<\frac{\mu}{2}$), with $b=a\frac{(n-1)!}{\alpha^(n-1)}$. Indeed, 
from Eq.~\eqref{specdim} we get that $|R_{l}(i)|\leq al^{n-1}\leq 
a\frac{(n-1)!}{\alpha^{(n-1)}}e^{\alpha l}$ for every $\alpha>0$. 
\end{cor}

Finally, we note that a bound can be valid even for more ``exotic'' spaces. If $\Gamma$ is a rooted 
tree graph with $n>1$ branches, we have that $|R_l(i)|= n^l+1$, so its fractal dimension 
is infinite. But we still have the following.
\begin{cor}
In Theorem~\ref{thm}, if $\Gamma$ is a rooted tree graph with $n$ branches, inequality~\eqref{cota} 
holds for $\mu\geq 2\ln n$, $\alpha=\ln n$, and $b=2$. Since $|R_{l}(i)|\leq 
2{n^{l}}=2e^{l\ln{n}}$ we just have to set $\alpha=\ln{n}$ and 
$b=2$.
\end{cor}

\section{Discussion}\label{conclusion}

First of all, let us explain in what sense we claim that entanglement 
satisfies an ``effective light-cone".
Let $\tilde{X}_{x}=\bigcup_{l=0}^{x}X_{l}$ be the enlargement of a subset $X\subset 
\Lambda$ up to distance 
$x$, as depicted in Figure~\ref{fig1}. Again, as in Sec.~\ref{results}, take 
$\rho^{q}_{\tilde{X}_{x}}(t)$ to be the reduced state in region $\tilde{X}_{x}$ of the evolved 
states $\ket{\psi^{q}(t)}$, where $q=0,1$, or $2$. Recall that the system evolves without 
perturbations if $q=0$ but is subjected to a local quench in region $X$, at $t=0$, in the 
Hamiltonian for $q=1$ or in the initial state if $q=2$. Let 
$E_{q}(x,t)=S(\rho^{q}_{\tilde{X}_{x}}(t))=S(\rho^{q}_{\Lambda-\tilde{X}_{x}}(t))$ be the entropy 
of entanglement of the evolved state $\ket{\psi^{q}(t)}$ under the bipartition defined by 
$\tilde{X}_{x}$, that is, $\Lambda=\tilde{X}_{x}\bigcup (\Lambda-\tilde{X}_{x})$. For a large class 
of models our results show that this entanglement function satisfies an \emph{effective} 
``light cone'', whatever the size $|\Lambda|$ of the whole system. Namely, by using 
inequality~\eqref{cota} with $Y=\Lambda-\tilde{X}_{x}$, we see that whenever 
$d(X,\Lambda-\tilde{X}_{x})=x\gtrsim v_{\mu}'t$ we 
shall have $|E^{0}(x,t)-E^{q}(x,t)|\approx 0$ for $q=1$ and $2$. Therefore, significant 
variations of entanglement can take place only inside the ``light cone'' $x\leq v_{\mu}'t$.

As a particular case of the above discussion, we point out some implications for 
$t-$DMRG  simulations of local quenches on spin chains. In such algorithms one 
naturally computes the entanglement of the system for every bipartition (in two 
contiguous regions) and every instant of time. These values for entanglement are 
important to establish how large the sizes of the matrices involved in the simulation must be in 
order to achieve good approximations.  In particular, a condition for the efficiency of the 
algorithms is that the simulated states must satisfy an area law for entanglement~\cite{arealaw}. 
Now, assume that a quench in the Hamiltonian is applied 
on an extreme point of the chain and take $x$ to be the distance between this site and the cutting 
point of a bipartition. As a particular case of the above discussion, we guarantee that 
$|E_{0}(x,t)-E_{1}(x,t)|$ satisfies the bound~\eqref{cota} with $x=d(X,Y)$ and has no dependence 
whatsoever with the size of the regions or the whole system. Now, if the initial state is an 
eigenstate of the unperturbed Hamiltonian and satisfies an area law, we have 
$E_{0}(x,t)=E_{0}(x,0)\leq c_{0}$, where $c_{0}$ is some constant. 
Therefore, by our bound, the evolved state will still satisfy an area law for any finite time.
Indeed, from Theorem \ref{thm} we have that $E_{1}(x,t)\leq c_0+c_1e^{\mu v|t|}$, for 
every $x\geq x_0$, where $c_1=\gamma_1 e^{-\mu x_0}$, and some fixed $x_0>v_{\mu}t$. Then, for some 
fixed value of $t$, we have an area law. Note that an area law, by itself, can 
already be drawn, for instance, from reference \cite{bounded}. There, the authors find that 
$E_{1}(x,t)\leq c_0+c'_1|t|$ holds for every $x$, where $c'_1$ is a constant dependent only on the 
parameters of the Hamiltonian. Our bound, however, can impose a stronger restriction on the 
entanglement growth for fixed $t$ and increasing values of $x$.

In Ref.~\cite{Bravyi} the authors show that a Lieb-Robinson bound indeed implies a limitation 
for the propagation of information throughout the many-body system in the information-theoretical 
sense. Assume two observers $A$ and $B$ have access to regions $X$ and $Y$ of the 
many-body system, respectively. They can establish a communication channel from $A$ to $B$ in the 
following way. Observer $A$ can encode an alphabet with $m$ letters in the state of the system by 
applying one out of $m$ unitaries on the initial state $\ket{\psi}$, all of them with support on 
region $X$. Observer $B$ can then perform measurements on region $Y$ in order to discern which 
unitary was applied and, hence, which letter of the alphabet was intended to be sent.  If $p_{i}$ is 
the probability for the $i$th letter to be sent, the maximum amount of information 
that can pass through this channel is measured by the Holevo capacity, given 
by $C(t)=S(\sum_{i=1}^{m}p_{i}\rho_{Y,i}(t))-\sum_{i=1}^{m}p_{i}(S(\rho_{Y,i}(t))$, where 
$\rho_{Y,i}(t)$ is the reduced state on $Y$ given by the evolution of $U_{i}\ket{\psi}$ at time 
$t$. 

Through a bound for $|S(\rho_{Y,i}(t))-S(\rho_{Y,j}(t))|$, for any $i\neq j$, the authors of \cite{Bravyi} show that the 
Holevo capacity is small for small times ($t\ll d(X,Y)/v_{\mu}$). Their bound, however, is
proportional to the volume of $Y$. Therefore, it is necessary to additionally assume this volume 
grows at most polynomially with $d(X,Y)$. By Example~\ref{dimn}, for systems with 
$n$ spatial dimensions, however, such additional assumption is no longer required. Even if observer 
$B$ has access to an arbitrarily large portion of the system, no significant amount of information 
can be sent through the channel for small times. 

We may add that the communication channel could be alternatively implemented by observer 
$A$ encoding the letters of the alphabet on Hamiltonian perturbations $W_{i}$ with support on 
$X$. Our results also guarantee the Holevo capacity would be small for small times, even for 
arbitrarily large regions $Y$.

\vspace{1.5cm}

\textit{Acknowledgements}.

We acknowledge financial support from Conselho Nacional de Desenvolvimento Cient\'ifico e Tecnol\'ogico (CNPq) and Coordena\c{c}\~ao de Aperfei\c{c}oamento de Pessoal de N\'ivel Superior (CAPES). We thank Fernando G. S. L. Brand\~ao and Alberto L. de Paula, Jr. for useful discussions.

\end{document}